# Broadband optical-fiber-compatible photodetector based on a graphene-$MoS_2$-$WS_2$ heterostructure with a synergetic photo-generating mechanism

Yi-feng Xiong, Jin-hui Chen and Fei Xu*
*feixu@nju.edu.cn

National Laboratory of Solid State Microstructures, College of Engineering and Applied Sciences and Collaborative Innovation Center of Advanced Microstructures, Nanjing University, Nanjing 210093, P. R. China

**ABSTRACT**

Integrating two-dimensional (2D) crystals into optical fibers can grant them optoelectronic properties and extend their range of applications. However, our ability to produce complicated structures is limited by the challenges of chemical vapor deposition (CVD) manufacturing. Here, we successfully demonstrate a 2D-material heterostructure created on a fiber endface by integrating a microscale multilayer graphene-$MoS_2$-$WS_2$ van der Waals (vdW) heterostructure film on it. Hence, based on simple layer-by-layer transferring, a visible-to-infrared gating-free all-in-fiber photodetector (FPD) is produced. Our FPD exhibits an ultrahigh photoresponsivity of ~$6.6×10^7$ A·$W^{-1}$ and a relatively fast response speed of ~ 7 ms at 400 nm light wavelength, due to the strong light absorption and the built-in electric field of the heterostructure. Moreover, owning to the type-II staggered band alignments in the $MoS_2$-$WS_2$ heterostructure, the interlayer optical transition between the $MoS_2$ and $WS_2$ layers enable our FPD to sense the infrared light power, displaying a photoresponsivity of ~ 17.1 A·$W^{-1}$ at 1550 nm. In addition, an inverse photoresponse is observed under an illuminating power higher than ~1 mW at 1550 nm, indicating a competingphotocurrent generation mechanism, comprising the photoconductive and photo-bolometric effects. These findings offer a new strategy for developing broadband and gating-free photodetectors and other novel optoelectronic devices based on 2D crystals. Furthermore, our fabrication method may provide a new platform for the integration of optical fibers with semiconducting materials.

The invention of optical fibers has contributed to the rapid development of communication, sensor, laser, and numerous other technologies.[1] However, due to the limitations of their material characteristics (most often $SiO_2$), optical fibers do not possess optoelectronic functions, which limits their applications. Integrating two-dimensional crystals serving as optoelectronic materials into optical fibers is an effective way to solve this problem. [2-4]

Along with the discovery of graphene, other two-dimensional (2D) materials—including transition-metal (Mo,W) dichalcogenides (TMDCs) —have received considerable attention for their potential applications in future electronic and optoelectronic devices (e.g., transistors, photodetectors, sensors, and solar cells).[5-10] However, for photodetection applications, individual 2D materials do not possess sufficiently high responsivity ($R$), fast response time ($\tau$), and broadband working bandwidth simultaneously. For example, photodetectors based on gapless pristine graphene display very fast time response and broadband detection spectra, but their responsivity is limited to a level of $mA·W^{-1}$, due to the low light absorption rate and fast electron-hole recombination in graphene. [8,11-13] Photodetectors based on molybdenum disulfide ($MoS_2$) show a high photoresponsivity of ~ 880 $A·W^{-1}$, but their response time is relatively long due to the long life time of their photocarriers, and their working wavelength is limited to the visible spectrum due to the large band gap of $MoS_2$.[14,15]

Considering the high carrier mobility inside graphene[16,17], an effective method of building high-performance photodetectors is to interface graphene with light-absorbing materials, such as quantum dots[18], perovskites[19], TMDCs[20,21], organic heterostructure[22,23], silicon[24], etc. Among these graphene-based van der Waals (vdW) heterostructures, photodetectors consisting of only atomically thin two-dimensional (2D) materials are particularly attractive for their possibility to form a sandwich structure for ultra-flexible and high-performance electronic and optoelectronic applications.[25] However, these types of graphene-based devices consisting of only one layer of 2D crystals also suffer from the trade-off between photoresponsivity ($R$) and response time ($\tau$). For example, graphene-$MoS_2$ heterostructure based photodetectors show enhanced photoresponsivity (~ $5×10^8$ $A·W^{-1}$) at the expense of extremely slow response times (~ $1×10^3$ s).[20] The absence of an internal electric field within the light-absorption layer can cause low photocarrier-separation efficiency, thus resulting in low quantum efficiency ($QE$) and relatively long response times (typically in the range of seconds, without gating pulses). [20] Although several strategies (e.g., the use of a ferroelectric substrate[26] or a top transparent electrode[18]) artificially introduce an external electric field to facilitate charge separation, these approaches are high-cost and have require complex fabrication processes. Furthermore, most of these photodetectors display a working wavelength range only in the visible spectra and cease to be effective for infrared wavelengths, including the communication wavelengths in silica fiber (e.g., 1310 nm, 1550 nm), owning to the limitation by the intrinsic band gap of the constituting materials. [27,28]

Herein, we successfully fabricate an all-in-fiber photodetector (FPD) by assembling a micrometer-scale multilayer graphene-$MoS_2$-$WS_2$ vdW heterostructure film with gold (Au) electrodes on a fiber endface, using a layer-by-layer transferring method.[29] We exploit the built-in field at the interfaces of the constituting layers to accelerate the separation of the electron-hole pairs instead of using the gating method to obtain an external electric field. Moreover, due to the difference in electronic band structures between the $MoS_2$ and $WS_2$, their type-II staggered band alignment can generate indirect optical excitons and induce charge spatial separation, thus decreasing the threshold energy of incident photons and enable our FPD to function as a broadband device from the visible to the infrared spectra.[25,30] Our gating-free FPD device exhibits an ultrahigh photoresponsivity of ~ $6.6×10^7$ $A·W^{-1}$ for an incident light with a wavelength of 400 nm, and ~ 17.1 $A·W^{-1}$ for a wavelength of 1550 nm. Furthermore, this device also shows a relatively high response speed of ~ 7 ms. Generally, an objective lens has to be used to focus the light source into a sub-sample-size scale to measure the performance of photodetectors on silicon substrates.[31] Theoretically, thanks to our all-in-fiber structure, we can easily concentrate all the input power into the FPD without any lens system.[32] We believe that this finding would provide a novel approach to designing broadband and gating-free photodetectors and other novel optoelectronic devices based on 2D materials, and our unique fabrication method may provide a new platform for the integration of optical fibers with semiconducting materials.

## RESULTS
**Device fabrication and characterization**

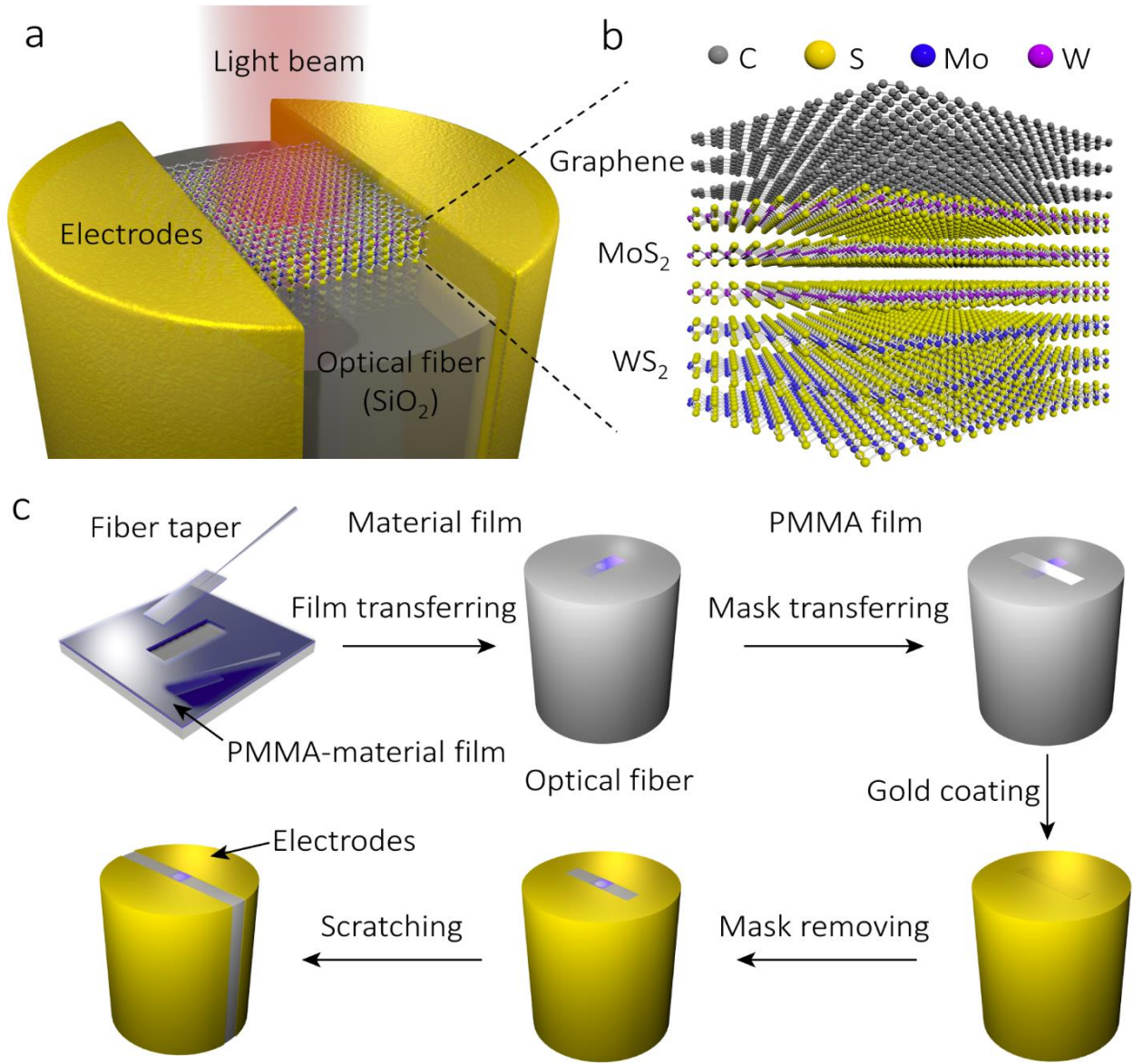

**Figure 1 Schematic of the FPD and its sequential fabrication process.** (a) 3D schematic view of the FPD illuminated with a monochromatic light beam. (b) Schematic structure of the graphene-MoS$_2$-WS$_2$ vdW heterostructure inside the FPD. (c) Sequential fabrication process of the FPD.

**Figure 1** shows the sequential fabrication process of the FPD: (1) the commercially available CVD two-dimensional material (graphene/MoS$_2$/WS$_2$) film (Six Carbon Technology Inc.) is transferred onto a SiO$_2$ substrate by coating a layer of polymethyl methacrylate (PMMA) and dissolving the original substrate (Cu/sapphire). (2) The PMMA-2D material film is cut into small rectangular stripes (e.g., 20 μm in width, 50 μm in length) and lifted under an optical microscope by using a fiber taper placed on a high precision three-dimension moving stage (NanoMax, Thorlabs). [29] (3) One of the PMMA-2D material stripes is transferred onto a cleaved single-mode optical fiber to cover its core. (4) The sample is baked at 180 °C for 20 minutes to improve the contact between the composite PMMA film and optical fiber substrate. After that, the PMMA layer is removed by immersing it in acetone for 5 minutes. To thoroughly remove the PMMA, the baking and immersion processes are repeated for several times. Sequentially, the

WS$_2$, MoS$_2$, and graphene films are transferred to an optical fiber endface step-by-step. (5) A stripe of PMMA serving as a mask is placed vertically onto the three films. (6) Physical vapor deposition (PVD) is used to deposit a ~40 nm thick gold film over the whole fiber facet. (7) The PMMA mask is removed by using a fiber taper. (8) A portion of the gold layer on the endface and the lateral wall of the fiber is scratched to obtain a small electrode channel.

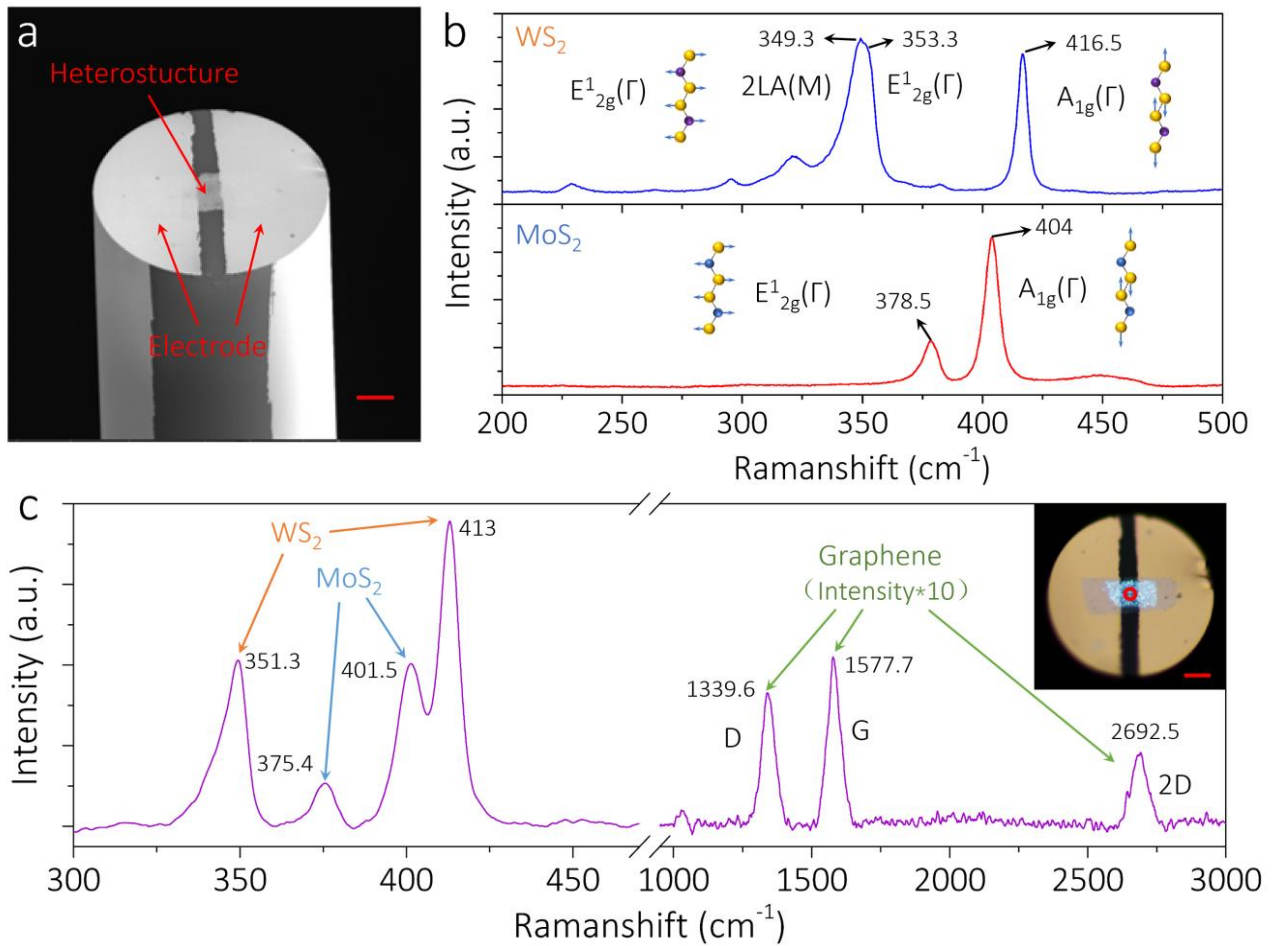

Figure 2 Characterization of the FPD. (a) SEM image of the FPD. The scale bar is 20 μm. (b) In situ Raman spectra of the multilayer WS$_2$ (top) and multilayer MoS$_2$ (bottom) transferred onto the silicon substrate, using a 532 nm laser. Inset: schematic diagram of the in-plane phonon mode E$^1_{2g}$ (Γ) and the out-of-plane phonon mode A$_{1g}$ (Γ). (c) In situ Raman spectra of the stacked multilayer graphene-MoS$_2$-WS$_2$ vdW heterostructure in the FPD, using a 532 nm laser. Inset: the optical microscope image of the stacked multilayer graphene-MoS$_2$-WS$_2$ vdW heterostructure on the fiber endface, where the red circle is the selected area for the Raman characterization. The scale bar is 20 μm.

**Figure 2a** shows a SEM image of the stacked graphene-WS$_2$-MoS$_2$ heterostructure integrated optical fiber structure. Two collecting electrodes are placed adjacent to the fiber core, and the width of the central channel approaches the diameter of the fiber core (~9 μm), which is beneficial for efficient photodetection. **Figure 2b** shows the Raman spectra of the CVD multilayer WS$_2$ and multilayer MoS$_2$ transferred onto the silicon substrate. Two typical optical-phonon peaks are observed in the Raman spectra of WS$_2$/ MoS$_2$: E$^1_{2g}$ (Γ) represents the in-plane optical mode, while A$_{1g}$ (Γ) corresponds to the out-of-plane vibrations of the sulfur atoms (inset of Figure 2b).[33] **Figure 2c** shows the Raman spectra of the transferred graphene-MoS$_2$-WS$_2$ heterostructure on the FPD (red circle in the inset of Figure 2c). The low-frequency region of the spectra contains two typical vibration modes of both WS$_2$ and MoS$_2$, indicating the formation of a heterostructure. The E$^1_{2g}$ (Γ) and A$_{1g}$ (Γ) modes are identified at 351.3 and 413 cm$^{-1}$ for WS$_2$, and 375.4 and 401.5 cm$^{-1}$ for MoS$_2$, respectively. At the high-frequency region, the three characteristic peaks for graphene at 1339.6 cm$^{-1}$ (D-band), 1577.7 cm$^{-1}$ (G-band), and 2692.5 cm$^{-1}$ (2D-band) are observed.

**Photodetection performance and analysis**

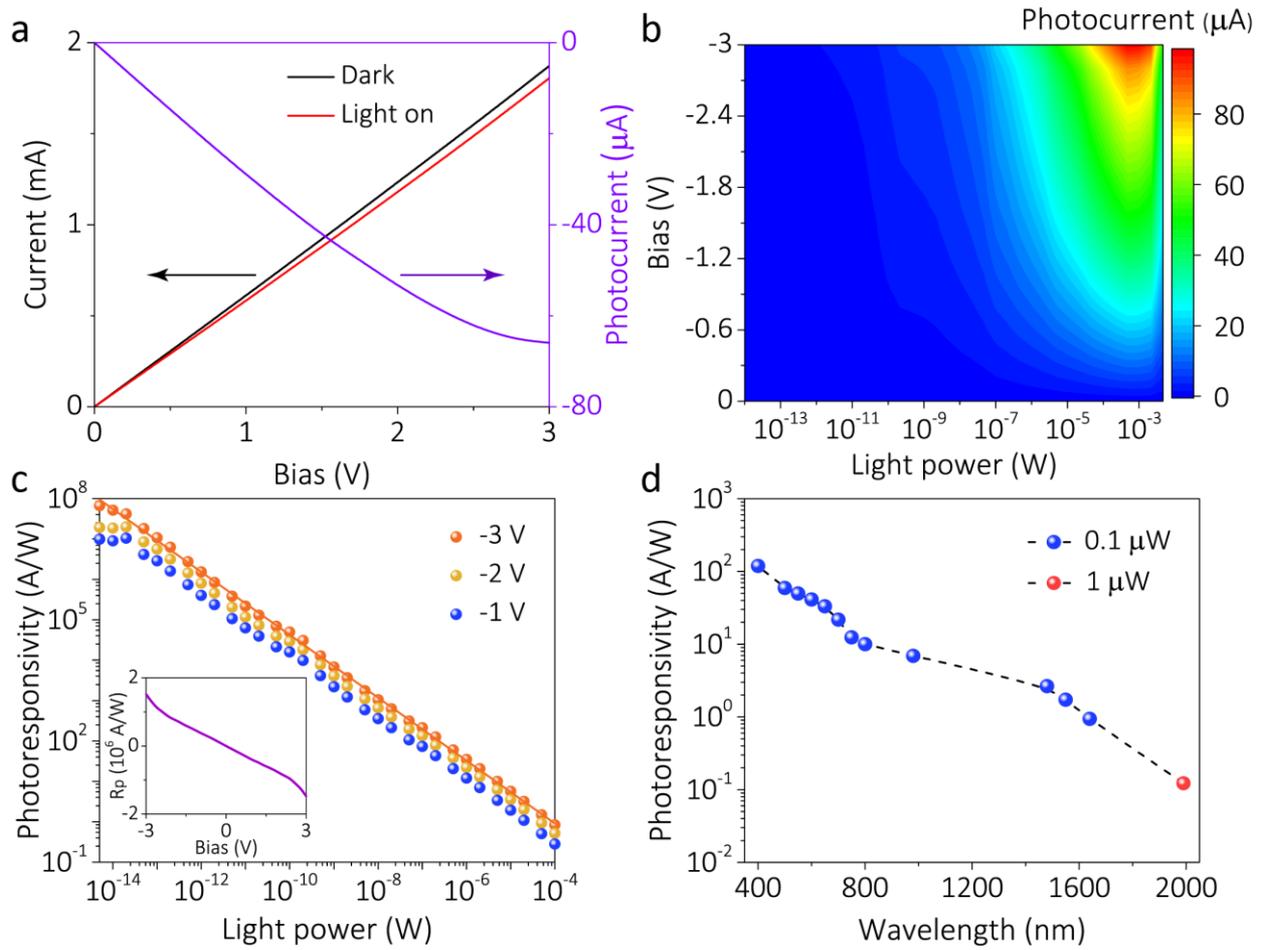

Figure 3 Photoelectrical characterization of the FPD under the illuminating wavelength of 400 nm. (a) Black and red lines are the I-V curve of the FPD in a dark environment and under illumination (0.1 mW), respectively. The violet line is the generated photocurrent under the illumination. (b) Photocurrent generation is related to the illuminating power and bias voltage. (c) Photoresponsivity of the FPD as a function of illuminating power and bias voltage. The red line is the fitting curve with the bias of -3 V. Inset: the photoresponsivity of the FPD is related to the bias voltage with a fixed input light power of 1 pW. (d) Absolute value of the photoresponsivity of the FPD as a function of the incident light wavelength. The incident light power is fixed at 100 nW from 400 nm to 1640 nm wavelength (blue dots) and 1 μW for 2 μm nm wavelength (red dot).

The I-V characteristic curve of the FPD is displayed in **Figure 3a**. The photocurrent ($I_p = I_{light} - I_{dark}$) is defined as the change of drain-source current with and without illumination (i.e., being placed in a dark environment). The linearity of the drain-source I-V curve suggests an Ohmic contact between the Au electrodes and the heterostructure. Ordinarily, photodetectors have an increasing conductivity under illumination due to the creation of photocarriers, while for our fabricated FPD device, the conductivity is reduced under illumination by a 400 nm light source (MDL-III-400, CNI, China). This indicates a special photogeneration mechanism. Moreover, the absolute value of the photocurrent is proportional to the drain voltage under illumination due to the increase of the carrier drift velocity. **Figure 3b** shows the photocurrent mapping relating the bias voltage and incident light power. At an incident light power below ~1 mW, the photocurrent increases with the bias voltage and light power. When the illuminating power is higher than ~1 mW, the photocurrent decreases, implying a synergetic photo-generating mechanism will be discussed in the next section. The photoresponsivity is one of the most important parameters of a photodetector, which is defined as

$$R = I_p / P_{light}$$

where $P_{light}$ is the incident light power. The near-linear curve (in a log-log plot) of the photoresponsivity of the FPD relating to the input light and the bias voltage is shown in Figure 3c, indicating that the trap-state of MoS$_2$ and WS$_2$ has a considerable influence on the FPD. The trap-state can greatly enhance the photoresponsivity of the FPD by increasing the carriers' life time and contributing to the photogating effect. With the increase of the illuminating light power, the trap-states are filled, leading to the decrease of photoresponsivity.[15] The inset of **Figure 3c** shows the photoresponsivity as a function of the bias voltage at a fixed light power of 1 pW, showing that the absolute value of the photoresponsivity increases with the increase of the bias voltage. The maximum photoresponsivity of the presented FPD device reaches ~ 6.64×10$^7$ A·W$^{-1}$ with a light intensity of ~5 fW (6.35 nW/cm$^2$) and a bias voltage of −3 V. The external quantum efficiency (EQE) is the ratio of the number of photo-generate charge carriers and the total number of impinging photons, which is closely related to the photoresponsivity:

$$EQE=R(hc/e\lambda)$$

where $h$ is Planck's constant, $c$ is light speed in a vacuum, $e$ is the quantity of electric charge, and $\lambda$ is the light wavelength. The EQE of the FPD reaches ~ 2.06×10$^8$ for a ~ 5 fW light input. Comparing to state-of-the-art graphene hybrid photodetectors, our FPD device exhibits a relatively high photoresponsivity and EQE considering gating-free conditions and simple device construction. In order to investigate the detecting light wavelength range of the FPD device, photoresponsivity was measured as a function of the incident light wavelength and is displayed in **Figure 3d**, where the incident light power was 100 nW for the illuminating wavelengths from 400 nm to 1640 nm and 1 μW for the wavelength of 2 μm. The device shows a decreasing photoresponsivity as the illuminating wavelength is increased from 400 nm to 2 μm, which indicates that the FPD device can be used for broadband photodetection, covering the visible and infrared ranges, while maintaining a high photoresponsivity. For ordinary MoS$_2$ or WS$_2$ devices, the photoresponse is usually negligible at wavelengths over ~ 680 nm (corresponding to the photon energy of ~ 1.8 eV), due to their large band gaps.[14] In the wavelength range of 400 nm to 1640 nm, our FPD device shows a similar photogeneration mechanism of photoconductive effect, while at the wavelength of ~ 2 μm, its conductivity increases under illumination, corresponding to the photo-bolometric effect.[15]

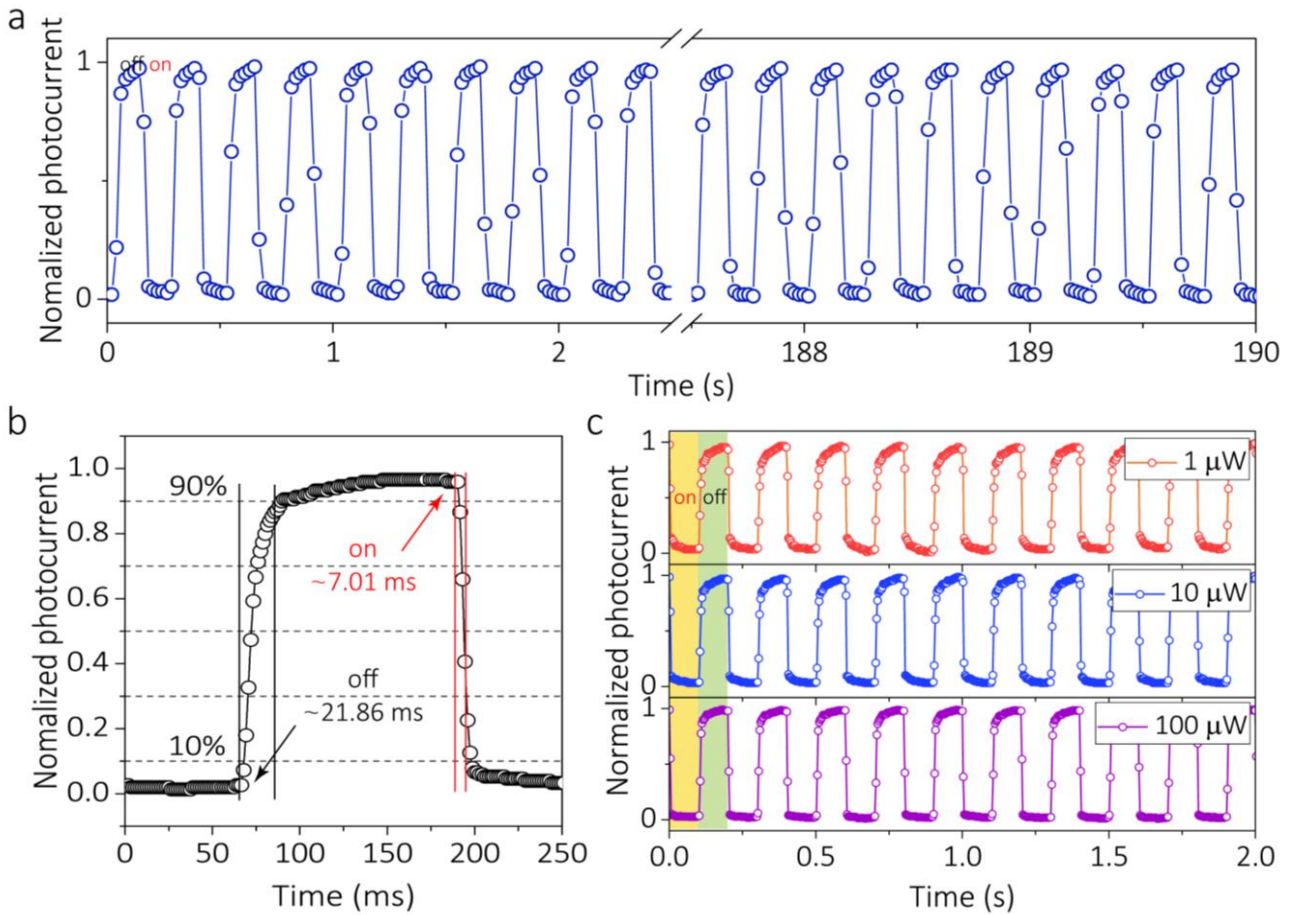

**Figure 4 Photocurrent dynamics and stability characterization.** (a) Stability of the FPD for over 750 cycles of light modulation (@400 nm), measured at 4 Hz, 1 V, and 100 μW. (b) View of the normalized photocurrent dynamics during 1 cycle of light modulation of the FPD, measured at 1 V and 100 μW. (c) Normalized photocurrent dynamics measured at 4 Hz and 1 V for several input light powers.

To further confirm the photodetection characteristics of the FPD, the drain-source current as a function of time under a periodically switched light source is shown in **Figure 4**. The current drops under illumination and recovers when the light is turned off, which illustrates a photogeneration mechanism different from ordinary photodetectors. Here, we repeat the ON-OFF cycles of the input light for over 750 times, as shown in **Figure 4a**, implying that the FPD photodetection exhibits a great stability. The response speed of the FPD is relatively fast, reaching up to 21.86 ms and 7.01 ms for the 10–90 % rise/fall time, respectively, as shown in **Figure 4b**. Moreover, the photocurrent dynamics of the FPD under three different illumination intensities are displayed in **Figure 4c**, which are almost identical.

**DISCUSSION**

Recognizing the working principle of the FPD is crucial for improving its performance. The photodetection performance parameters of FPDs based on the multilayer graphene, a $MoS_2$-$WS_2$ heterostructure, and graphene-$MoS_2$-$WS_2$ heterostructure are measured respectively and compared in **Table 1**. The graphene layer not only plays a role of the conductive layer, but also participates in the photosensing, resulting in a positive photocurrent based on the photo-bolometric effect with a light power higher than ~0.1 mW. The photoresponsivity of the $MoS_2$-$WS_2$ heterostructure is only in the order of tens of mA·W$^{-1}$, due to its low carrier mobility and relatively large electrode gap. Hence, compared to the graphene and $MoS_2$-$WS_2$ heterostructure, the graphene-$MoS_2$-$WS_2$ heterostructure has a significantly better performance. Moreover, the sign of the photocurrent in the graphene-$MoS_2$-$WS_2$ heterostructure is inverted–unlike the other structures–indicating that a novel photodetection mechanism is involved.

**Table 1** Photodetection performance parameters of FPDs based on the multilayer graphene, $MoS_2$-$WS_2$ heterostructure, and graphene-$MoS_2$-$WS_2$ heterostructure.

| Parameters | graphene | $MoS_2$-$WS_2$ | graphene-$MoS_2$-$WS_2$ |
|---|---|---|---|
| Photogenerate mechanism | bolometric | photoconductive | photoconductive bolometric |
| Sign of photocurrent | positive | positive | negative |
| Detection limit | 0.1 mW (1550 nm) | 0.2 nW (400 nm) | 5 fW (400 nm) / 20 nW (1550 nm) |
| Responsivity(A/W) | $2.86 \times 10^{-3}$ | $8.49 \times 10^{-2}$ | $6.6 \times 10^{7}$ (400 nm) / 17.1 (1550 nm) |
| EQE | $2.29 \times 10^{-3}$ | 0.264 | $2.06 \times 10^{8}$ (400 nm) / 13.71 (1550 nm) |
| Time response | ~100 ms | ~10 ms | ~7 ms (200 μW) / ~160 ms (5 mW) |
| Wavelength range | visiable and infrared | visiable | Visible and infrared |

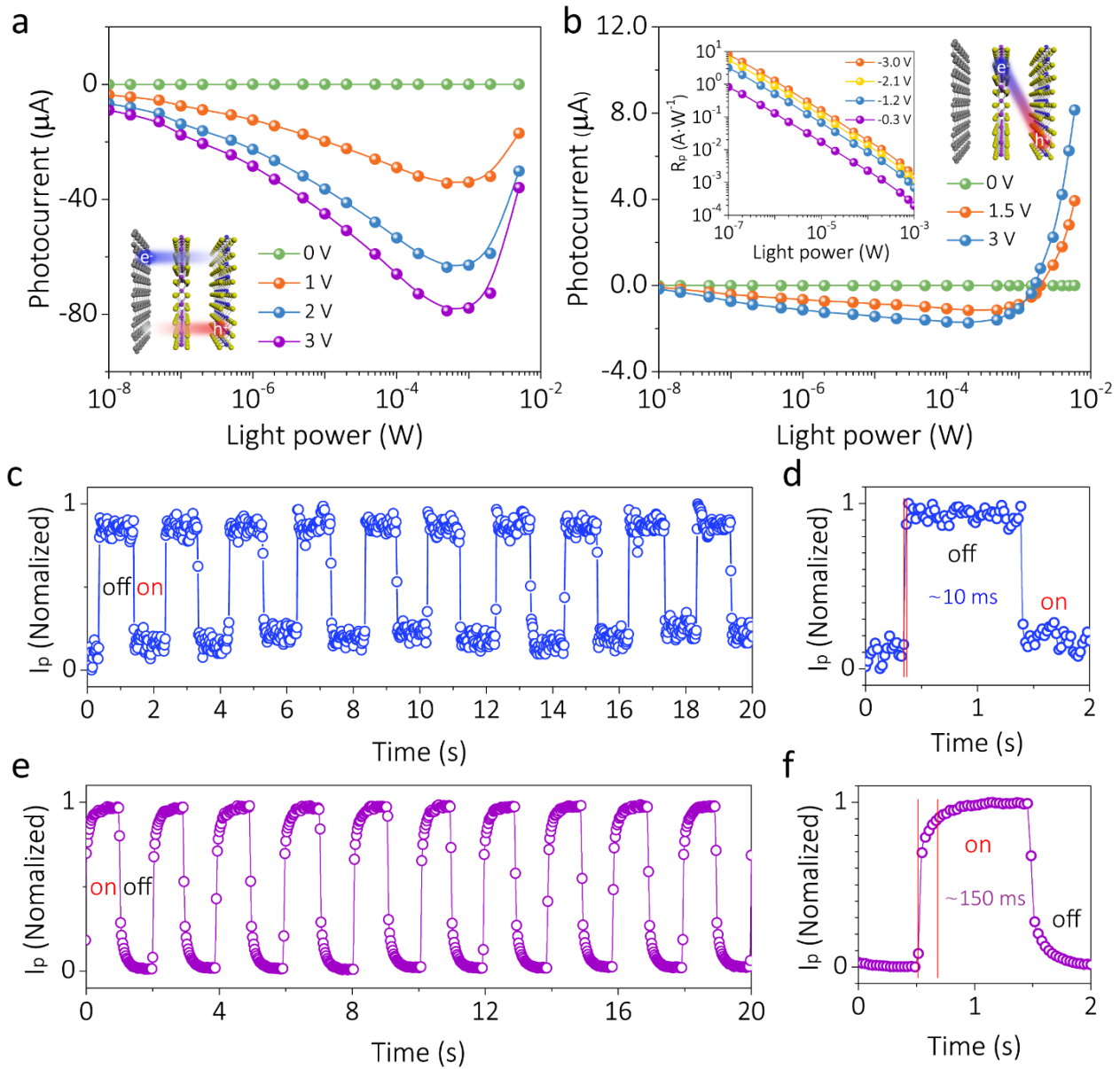

Figure 5 Photoresponse characteristics of the FPD. (a) Photocurrent generation as a function of illuminating power and bias voltage with an input wavelength of 400 nm and (b) with an input wavelength of 1550 nm. Inset of (a): direct excitons in the the $MoS_2$-$WS_2$ heterostructure. Inset of (b), left: the photoresponsivity of the FPD as a function of illuminating power and bias voltage with an input wavelength of 1550 nm. Right: Indirect excitons in the $MoS_2$-$WS_2$ heterostructure. (c) Normalized photocurrent dynamics of the FPD, measured at a frequency of 0.5 Hz and bias voltage of 1 V. The incident light wavelength is 1550 nm and light power is 200 μW, while (e) the incident light power is 5 mW. (d) and (f) Normalized photocurrent dynamics during 1 cycle of light modulation in (c) and (e), respectively.

**Figure 5a** and **Figure 5b** show the photocurrent as a function of input light power with illumination wavelengths of 400 nm and 1550 nm, respectively. As was mentioned above, the graphene-$MoS_2$-$WS_2$ heterostructure generates a negative photocurrent under an illumination below ~ 0.1 mW. The absolute photocurrent increases with the increase of the incident power, due to the photoconductive effect. However, when the incident power is higher than ~ 0.1 mW, it increases due to the photo-bolometric effect of the graphene layer. As a result, the photocurrent produced by the photoconductive and photo-bolometric effects are opposite in sign and will cancel each other out. In **Figure 5a** (illumination wavelength of 400 nm), the negative photocurrent generated by the photoconductive effect is larger than the positive photocurrent generated by the photo-bolometric effect of the graphene layer, thus the whole photocurrent shows a decreasing tendency for light powers over ~ 1 mW. However, the sign of the photocurrent does not change. Meanwhile, under the illumination by a 1550 nm light source (TSL-710, Santec), the $MoS_2$-$WS_2$ layer generates fewer photocarriers due to the limited band gap, thus the positive photocurrent generated by the graphene layer reverses the sign of whole photocurrent under incident powers beyond ~1 mW, which is shown in **Figure 5b**. The current dynamics of the FPD are measured under different incident light powers: 200 μW (**Figure 5c**) and 5 mW (**Figure 5d**) with the same light wavelength of 1550 nm. As shown in Figure 5c, the current decreases under the illumination power of 200 μW, while in Figure 5d, the current increases under the illumination power of 5 mW, matching the mechanisms discussed above. In addition, the photoresponse times derived from the $MoS_2$-$WS_2$ and graphene layers are different due to their different photogeneration mechanisms, corresponding to ~10 ms and ~150 ms, respectively.

Herein, we propose the following photodetection mechanism to gain insight into the unique photoresponse behavior of our fabricated FPD based on a graphene-$MoS_2$-$WS_2$ vdW heterostructure. In the graphene-$MoS_2$-$WS_2$ heterostructure, a built-in electric field is formed due to the mismatch of the work functions of graphene, $MoS_2$, and $WS_2$. The direction of the electric field is from the graphene to the $MoS_2$ to the $WS_2$ layer. As a result, the photogenerated electron-hole pairs will separate automatically when light is focused on the heterostructure, and the electrons will be transferred from the $MoS_2$ and $WS_2$ to the graphene, while the holes will be transferred from the graphene and $MoS_2$ to the $WS_2$. Due to the PMMA residuals, the graphene is p-type[34], and hence, the electrons injected from the $MoS_2$ and $WS_2$ will decrease the concentration of its holes, leading to the rise of its Fermi level, thus decreasing its conductivity.[35] Moreover, the holes that are transferred and trapped in the $MoS_2$ and $WS_2$ layers can act as a local gate, providing a positive gate voltage to further decrease the conductivity of graphene. Therefore, both effects mentioned above can produce a negative photocurrent, resulting in a photosensitive behavior which is different from that of ordinary photodetectors.[27,36]

Interestingly, the cut-off wavelength of the individual $MoS_2$ or $WS_2$ is in the visible spectral range, yet, our FPD functions in the infrared wavelengths. In terms of the infrared light with photon energies below the bandgap of $MoS_2$ or $WS_2$, optical interband transition in the $MoS_2$ or $WS_2$ layers is prohibited, while the interlayer optical excitation and the charge spatial separation between the $MoS_2$ and $WS_2$ layers is allowed.[30] When infrared light is concentrated on the FPD, the photo-excited electrons from $WS_2$ are accumulated in $MoS_2$, while photo-excited holes from $MoS_2$ are accumulated in $WS_2$, which greatly reduces the photo-excitation band gap. Compared to the interband transition in the $MoS_2$ or $WS_2$ layers, the photogeneration efficiency of the interlayer transition between $MoS_2$ and $WS_2$ is much lower. As shown in the inset of **Figure 5b**, the maximum photoresponsivity of our FPD under the illuminating wavelength of 1550 nm reaches 17.1 A·W$^{-1}$ at the light intensity of ~ 20 nW (25.4 mW/cm$^2$), which is much larger than exhibited by state-of-art infrared photodetectors.

## CONCLUSIONS

In summary, we proposed and demonstrated a gating-free all-in-fiber photodetector (FPD) by integrating a micrometer-scale CVD multilayer graphene-$MoS_2$-$WS_2$ vdW heterostructure film onto a fiber endface. Our fabrication method is simple, low-cost, and repeatable. Thanks to the internal built-in field of the heterostructure, photogenerated electron–hole pairs can be effectively separated at the graphene-$MoS_2$-$WS_2$ interfaces, thus obtaining an ultrahigh photoresponsivity, wide bandwidth, and relatively fast response speed. Due to the interlayer optical excitation and the charge spatial separation between the $MoS_2$ and $WS_2$ layers, our FPD displays a wide spectrum response, including the infrared spectrum. Moreover, a synergetic photocurrent generation mechanism–including the photoconductive and photo-bolometric effects–was found in our FPD. We believe that our method may provide a new platform for the integration of optical fibers with semiconducting materials, and a new strategy for next-generation broadband electronic and optoelectronic devices.

## METHODS

**Test system and methods.**

We set up a test system including an optical circuit and an electrical circuit to measure the photosensing performance of the FPD. Since the FPD is naturally compatible with optical fiber systems, the incident light is transported by the optical fiber waveguide and used to illuminate the FPD directly. Here, we use a reference optical power-meter (S150C and S145C, Thorlabs) to calibrate the input light power. In photocurrent dynamics test, we use an optical chopper (Model C-995, Scitec) to realize the illumination ON-OFF switches. The generated electrical signal is collected and analyzed by a digital sourcemeter (Keithley SMU 2450, Tektronix).